# SPONTANEUOS AND PARAMETRIC PROCESSES IN WARM RUBIDIUM VAPOURS

M. Dąbrowski, M. Parniak, D. Pęcak, R. Chrapkiewicz, W. Wasilewski

Institute of Experimental Physics, Faculty of Physics,
University of Warsaw,
Pasteura 5, Warsaw 02-093, POLAND
e-mail: radekch@fuw.edu.pl

Warm rubidium vapours are known to be a versatile medium for a variety of experiments in atomic physics and quantum optics. Here we present experimental results on producing the frequency converted light for quantum applications based on spontaneous and stimulated processes in rubidium vapours. In particular, we study the efficiency of spontaneously initiated stimulated Raman scattering in the Λ-level configuration and conditions of generating the coherent blue light assisted by multi-photon transitions in the diamond-level configuration. Our results will be helpful in search for new types of interfaces between light and atomic quantum memories.

**Keywords:** *Raman scattering, rubidium vapours, quantum memory, four-wave mixing, spatial correlations, twin-beams, diamond configuration, coherent blue light generation.*

## 1. INTRODUCTION

Quantum properties of light and matter not only are the subject of fundamental research and studies but they are also becoming frequently applied in emerging quantum technologies. The most prominent and recognizable examples of applications of these technologies are the quantum cryptography [1], the quantum random numbers generators [2], and the quantum computation [3].

Quantum experiments are mostly carried out in two fields: the study of quantum properties of light and the study of the properties of ultracold atoms (mostly in Bose-Einstein's condensates). These two fields practically do not intersect, being focused on entirely different technological problems and scientific goals. On the edge of these fields, scientists run experiments using warm alkali metal vapours where the most popular medium is warm rubidium vapours. Rubidium atoms heated to temperatures ~ 100°C exhibit high optical depth and, as a consequence, provide high light to matter coupling [4].

Typically, experiments with warm rubidium vapours require a single heated sealed glass cell placed in magnetic shielding. The setups are much more robust, smaller and easier to maintain as compared with those designed for cold atoms. Two first rubidium optical transitions are on the wavelengths 780 nm and 795 nm, where commonly available Ti:Sapphire and diode lasers operate.

The vast range of experiments which have been run with warm rubidium vapours prove the versatility of this medium – in particular, as that for ultra-precise magnetometer measurements [5], where precision can even be enhanced by spin squeezing [6]. Atoms near resonance have highly nonlinear properties, which can also be used for generating the squeezed states of light [7].

The prominent and significant feature of atoms is their rich level structure enabling generation and storage of quantum superposition; consequently, atoms can serve as quantum memories [8].

This can be accomplished using large atomic ensembles [4], since the information can be stored in the collective excitation [9]. This is the only way to employ the warm vapours of the concentration of the order of $10^{12}$ cm$^{-3}$ and fast thermal motion where single atom addressing is impossible. This concentration is low enough to prevent atoms from frequent interactions. Further relevant techniques – such as the use of a buffer gas [10, 11] or paraffin coating of cell walls [12] – prolong the lifetime of these excitations from the range of microseconds [13] or milliseconds [14]; they even enable the generation of steady entangled states between two distant cells [15].

Atoms are used for storage of images [16], light pulses [17] or slowing the light [18] in the process of electromagnetically induced transparency [19].

From a vast array of processes employed for light-mater interaction we can single out the spontaneous and the stimulated processes. Such division is well known, as it appears in descriptions of the laser operation initiated by spontaneous emission and further amplified in the stimulated process. Other spontaneous and stimulated processes in nonlinear media provide a source for quantum light, e.g. the commonly used process of spontaneous parametric down-conversion [20].

In this paper we focus on two- and multi-photon spontaneous and stimulated processes related to the Raman scattering and frequency conversion with the use of rubidium atoms having rich level structure.

Our work is inspired by a recently proposed protocol of long distance quantum communication using warm atomic cells [21]. In this scheme, atomic vapours are used as quantum repeaters, with the mechanism of light-matter interaction based on collective Raman scattering.

Further motivation for our studies is the application (so far relatively weakly explored) of multi-spatial-mode properties of created and stored light, and, on the other hand, the usage of higher excited states of rubidium for quantum light generation. We aim at generation of quantum light and storage of quantum information. In this paper, we show two complementary approaches for generation and storage of quantum light based on spontaneous and stimulated collective processes.

## 2. THEORY

The starting point for describing in space and time the evolution of light and atomic excitation is to consider the interaction of a single atom with light. The base unit for our description is the Λ system, which in our case consists of three levels: two sublevels, $|g\rangle$ and $|h\rangle$, from the hyperfine manifold of atomic ground state, and an excited state $|e\rangle$. In the case of the $^{87}$Rb atom, $|g\rangle = |5^2S_{1/2}, F = 1\rangle$, $|h\rangle = |5^2S_{1/2}, F = 2\rangle$, and $|e\rangle = |5^2P_{1/2}\rangle$ as can be seen in Fig. 1$a,b$. Before the main part of the experiment atoms must be prepared in the $|g\rangle$ state. Writing information means transferring these atoms to state $|h\rangle$ via state $|e\rangle$ and collecting Stokes' photons from Raman scattering, whereas reading is connected with the opposite transition — from $|h\rangle$ to $|g\rangle$, in which Raman anti-Stokes photons with desired properties are produced. Raman scattering occurs when the Raman pump laser (write or read) is detuned far from the atomic resonance. The process of Raman scattering is quite complex, because writing and reading, i.e. Stokes and anti-Stokes scattering, are occurring simultaneously, with different amplitude probabilities. Moreover, real atoms consist of more than one $\Lambda$ system due to the hyperfine splitting and degeneracy, which makes theoretical predictions harder.

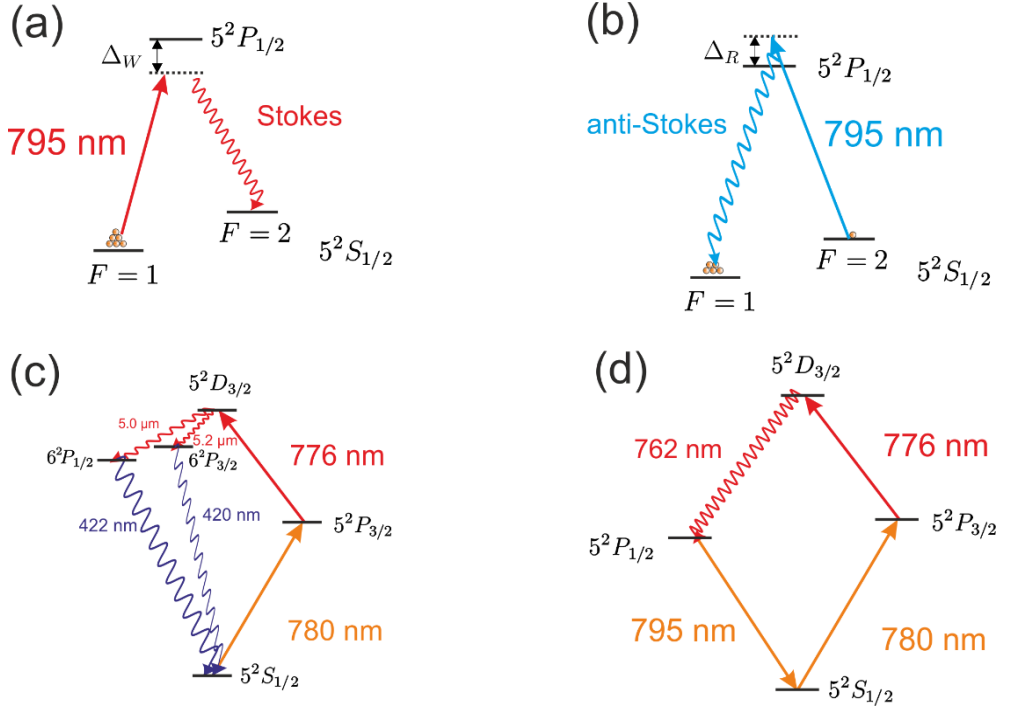

*Fig. 1*. Schemes of processes in $^{87}$Rb atom: a) Stokes scattering in the $\Lambda$ configuration., b) anti-Stokes scattering in the $\Lambda$ configuration, c) blue light generation in the diamond configuration, d) non-degenerate four-wave mixing in the diamond configuration.

We employ a semi-classical description of Raman scattering. This means that the light is represented classically by a vector of electric field:

$$E(r,t) = \frac{1}{2}\big(A_1(r)e^{i(-k_1 r - \tilde{\omega}_1 t)} + A_1^*(r)e^{i(-k_1 r - \tilde{\omega}_1 t)}\big) +$$
$$+ \frac{1}{2}\big(A_2(r)e^{i(-k_2 r - \tilde{\omega}_2 t)} + A_2^*(r)e^{i(-k_2 r - \tilde{\omega}_2 t)}\big). \tag{1}$$

As seen, there are two frequencies: $\tilde{\omega}_1$ and $\tilde{\omega}_2$, which represent Stokes and anti-Stokes photons with complex amplitudes $A_1$ and $A_2$, respectively.

The total Hamiltonian of a three-level atom in the $\Lambda$ configuration coupled with light is [22]:

$$\hat{H} = \begin{pmatrix} 0 & 0 & 0 \\ 0 & \hbar(\omega_1 - \omega_2) & 0 \\ 0 & 0 & \hbar\omega_1 \end{pmatrix} - \hat{d}E(r,t), \tag{2}$$

where $\omega_1$ and $\omega_2$ stand for natural frequencies of atomic transitions $|g\rangle \to |e\rangle$ and $|h\rangle \to |e\rangle$, and $\hat{d}$ is a dipole moment operator.

It is much easier to operate in the interaction picture, where Hamiltonian is time-independent. Applying it to our case and using the rotating wave approximation (i.e. omitting terms oscillating with double frequency), we obtain the following result:

$$\hat{H}_I = -\frac{\hbar}{2}\begin{pmatrix} 0 & 0 & \Omega_1^* \\ 0 & -2\delta & \Omega_2^* \\ \Omega_1 & \Omega_2 & 2\Delta \end{pmatrix}, \tag{3}$$

where $\Delta$ is the single photon detuning (e.g. $\Delta_W$ and $\Delta_R$ for writing or reading processes, respectively), $\delta$ is the two-photon detuning, and $\Omega_1$, $\Omega_2$ are Rabi frequencies.

The evolution including decoherence is described by the modified Heisenberg equation for the density matrix:

$$\frac{d\hat{\rho}}{dt} = -\frac{i}{\hbar}[\hat{H}, \hat{\rho}] - \frac{1}{2}(\hat{\rho}\hat{\Gamma} + \hat{\Gamma}\hat{\rho}), \tag{4}$$

where decoherence operator $\hat{\Gamma}$ takes the following form:

$$\hat{\Gamma} = \begin{pmatrix} \gamma & 0 & 0 \\ 0 & \gamma & 0 \\ 0 & 0 & \Gamma \end{pmatrix}. \tag{5}$$

Here $\gamma$ is the gross rate of ground state spin relaxation and the influx of thermalized atoms to the interaction region, while $\Gamma$ is the inverse lifetime of the excited state.

Equation (4) can be solved assuming that the population of highly excited levels could be omitted and using adiabatic elimination. We thus obtain the following [23]:

$$\frac{\partial B(z,t)}{\partial t} = c_\Delta A_1(z,t) - s_\Delta B(z,t) \tag{6}$$

$$\frac{\partial A_1(z,t)}{\partial z} = c_\Delta B(z,t), \tag{7}$$

where $B(z,t)$ is proportional to the coherence terms $\rho_{gh}$ of the density matrix $\hat{\rho}$. Including more excited levels in this model is relatively simple. Namely, the evolution equations are analogous but summing all excited states:

$$c_\Delta = -\sqrt{\frac{n\omega_1}{\hbar^2 c \varepsilon_0}} \sum_e \frac{A_2 d_1 d_{e2}^*}{2(\Gamma + 2i\Delta_{1e})\hbar}, \tag{8}$$

$$s_\Delta = \sum_e \frac{|A_2 d_{2e}|^2}{2(\Gamma + 2i\Delta_{1e})\hbar^2}. \tag{9}$$

It should be mentioned that $|C_\Delta|^2$ is proportional to the probability of the writing process. The real part of $s_\Delta$ can be interpreted as an inverse memory lifetime, which is also proportional to the intensity of light.

Omitting excited levels in the above approximation gives us an effectively two-level Hamiltonian that contains only the levels important for quantum memory processing – $|g\rangle$ and $|h\rangle$. This approach is adopted with success in the experiment – there was of course no need for considering the excited states in the storage process due to its short lifetime, but also during writing and reading the atom can be considered to be effectively only two-level.

Stokes scattering is a spontaneous stochastic process during which a photon and a correlated excitation are randomly generated in the medium. One cannot control what state is created, but after the Stokes photon is detected the information about the state and the spatial properties of excitation is gained, which can be used afterwards.

The quantum information stored in the ground-state atomic coherence can be subsequently retrieved by the deterministic process of anti-Stokes scattering.

The process of four-wave mixing depicted in Fig. 1c,d could also be described by considering the Hamiltonian of a multi-level atom coupled to optical fields. In the case of blue light generation (Fig. 1c) the highest excited state $5^2D_{3/2}$ is optically pumped, and a population inversion arises between this level and $6^2P_{1/2}$ or $6^2P_{3/2}$. Spontaneously emitted infrared photons are subject to strong amplification due to high dipole moment of the relevant transition. Blue light arises from resonant third-order nonlinearity leading to coherent four-wave mixing. The same holds for the second scheme (Fig. 1d), where we apply three optical fields and obtain a coherent emission at 762 nm due to the four-wave mixing process.

## 3. RAMAN SCATTERING: EXPERIMENTAL

The experimental setup we used is described in detail in [13]. For experiments we took a glass cell containing pure $^{87}$Rb isotope [24], with krypton as a buffer gas under the pressure of 0.5 Torr. The glass cell was heated to ~ 90ºC by bifilar magnetic coils. The cell was put into the $\mu$-metal magnetic shielding [25] to avoid decoherence produced by the external magnetic field.

During the experiment we used three external cavity diode lasers (ECDLs) at 795 nm: pump, write and read ones, working on the D1 line of $^{87}$Rb. Inside the cell the write and read lasers had the power of several mW, while the pump laser had the power of ~ 80 mW. Exact values of the detunings of write and read lasers were determined using a Doppler-free saturated absorption spectroscopy and their beat-note signal measured on a fast photodiode. All the lasers were locked using the the dichroic atomic vapour laser lock (DAVLL) setup [26]. Both writing and reading beams cross inside the rubidium cell at a small angle so that they superpose each other. The same crossing occurs between the abovementioned beams and the pumping beam. Pulses with duration from several to hundreds of microseconds were generated by acousto-optic modulators (AOM). The pulse sequence used in the experiment is shown in Fig. 2.

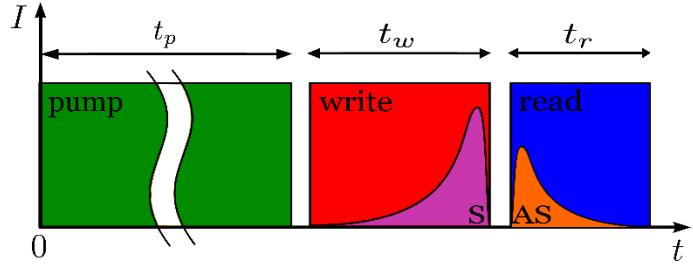

*Fig. 2*. Pulse sequence during the experiment: pumping, writing, storing and reading the quantum memory; typical temporal profiles of Stokes (S) and anti-Stokes (AS) pulses.

In the optical pumping process the pumping pulse with duration $t_p$=700 µs was used [27]. After 5 µs from the end of the pumping pulse we sent the writing pulse with variable duration $t_w$. The writing pulse light was scattered in the process of Stokes scattering inside the rubidium cell. The scattered light registered on the EM CCD (Electron Multiplying CCD, *Hamamatsu C9100-13*) camera indicated the presence of atomic coherence. Immediately after the writing pulse we sent the reading pulse of variable duration $t_r$. Due to the interaction between the reading pulse and the atomic coherence in the system, anti-Stokes light was produced. Once this light was registered on the camera, we received the proof that we had built a working scheme of quantum memory. The camera was situated in an optically far field, so the photons propagating in different directions were visible in different regions of the camera.

## 4. RAMAN SCATTERING: RESULTS

Our first task was to measure the efficiency of Stokes and anti-Stokes scattering as a function of duration and power of the writing laser. We measured signal from pixels in a 10×10 px region around the maximum value of each of the twin-beams [7] corresponding to an angle region of 760 µrad. The signal from the EM CCD was averaged over 1000 frames after the background signal had been subtracted. Laser powers were 9.2 mW and 3.5 mW for writing and reading beam, respectively. The read laser frequency was resonant to the $^{85}$Rb F=2 → F'=2 transition, whereas the write laser was red-detuned from the $^{87}$Rb F=1 → F'=1 resonant transition at $\Delta_W = 2\pi \times 2.508$ GHz [24]. The reading pulse duration was $t_r$=10 µs. Diameters of the beams were measured using the CCD camera: 3 mm, 2.5 mm and 6 mm for writing, reading and pumping laser beam, respectively. The results are shown in Fig. 3. The observed intensity of anti-Stokes light was eight times greater at increasing the writing pulse duration twice and four times greater upon increasing this power twice.

The theory considered in Sect. 2 predicts a linear scaling of the scattered light intensity with the write laser power (see Eq. 8), which agrees with measurements only in small regions of data points depicted in Fig. 3. Nevertheless, the trend of scattering efficiency growth with laser power corresponds well to the measured data. Thus, we were able to control the quantum memory efficiency by changing the writing laser properties.

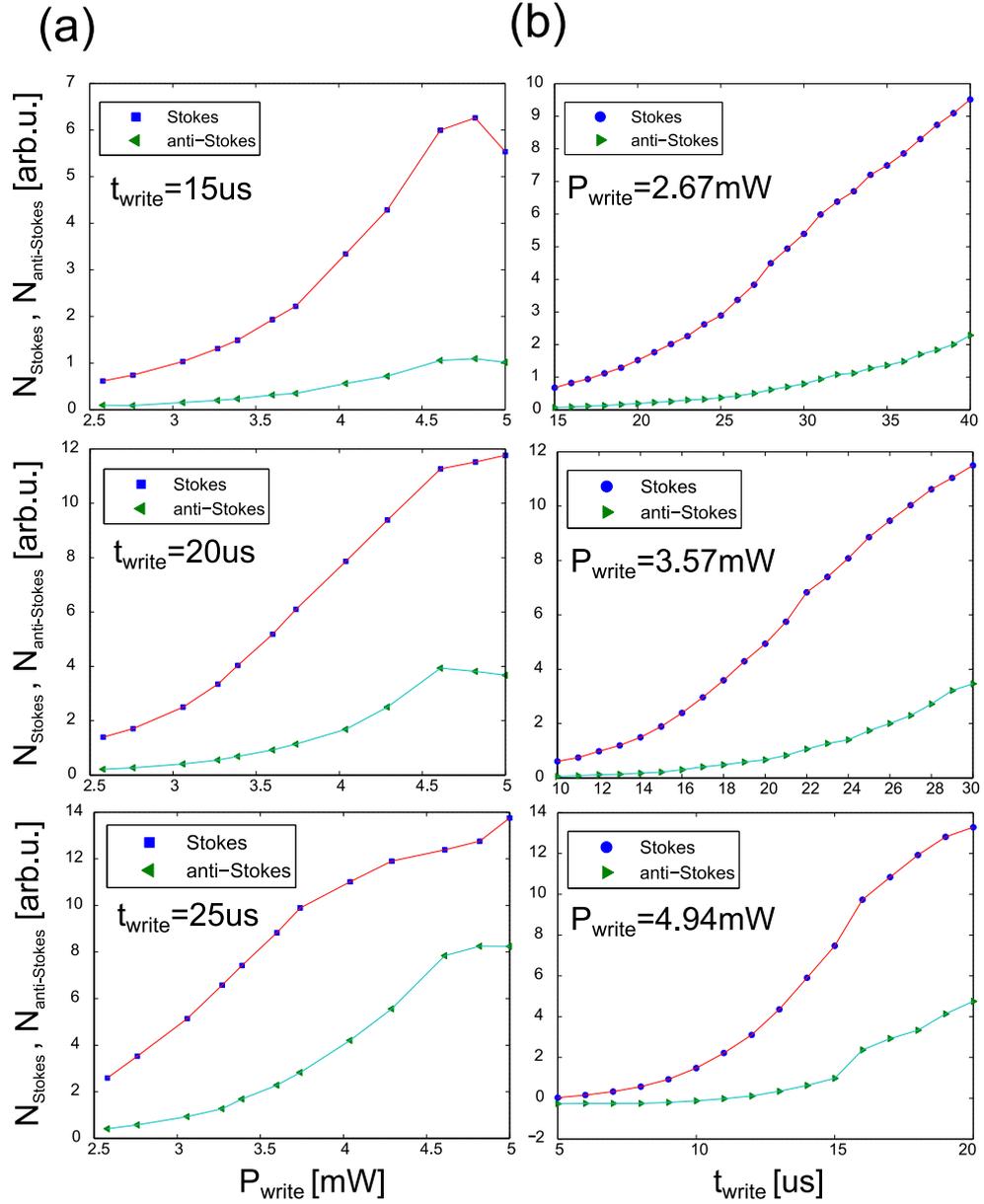

*Fig. 3*. Raman scattering beam intensity *vs.* duration of: writing pulse power (a); writing pulse (b) for different parameters.

The growth of Raman scattering signal was slower than the one measured in an analogous experiment [13]. We expected that this was due to the $\mu$-metal shielding remagnetization. To demagnetize the shielding we used a procedure proposed in [28]. The comparison of the signal growth before and after remagnetization procedure is shown in Fig. 4*a,b*. The experimental sequence of parameters after remagnetization was the same as before. We fitted the exponential curves $N_i(t) = N_i(0)e^{\alpha_i t}$ (where $i \in \{S, AS\}$) to the first 10 measured data points depicted in Fig. 4a,b and observed approximately a 30% growth of exponential coefficients $\alpha_S$, $\alpha_{AS}$ for both Stokes and anti-Stokes signals.

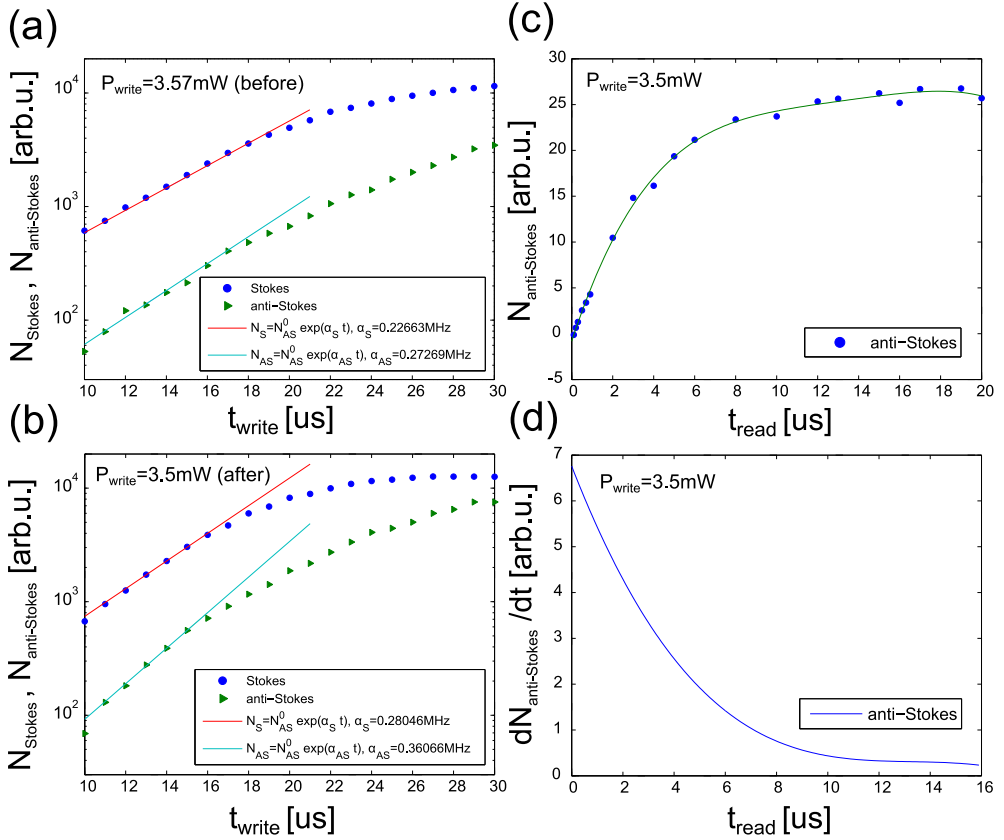

*Fig. 4.* Intensity of Raman scattering light before and after demagnetization (left). Anti-Stokes beam decay dynamics (right). Vertical plot scale is in arbitrary units.

Another issue studied in this experiment was the temporal dynamics of anti-Stokes light. In this case we changed the reading pulse duration $t_r$ within writing pulse duration $t_w$=20 $\mu$s that remained constant during the whole experimental sequence. By this we were able to probe the number of atomic excitations inside the rubidium cell as a function of time, since we assume that all of the atomic excitations were converted to anti-Stokes light during the readout process. The results are presented in Fig. 4c,d. We fitted the spline interpolation curve to the measured data and calculated the exponential decay curve $N_{AS}(t) = N_{AS}(0)e^{-\gamma t}$ by numerical differentiation. By repeating such a measurement procedure many times we obtained the exponential coefficient $\gamma = (200 \pm 100)$ kHz. Large spread of the obtained values is probably connected with fluctuations of laser power in time.

## 5. BLUE LIGHT GENERATION

In the following two sections we describe two simple experiments to demonstrate the usefulness of four-wave mixing processes using higher excited levels of a rubidium atom for generating the coherent states of light.

Although the coherent blue light generated in warm rubidium vapours has been the topic of significant interest among atomic physicists for some years now (beginning with the work by *Zibrov et al.* [29]), this is yet to be fully understood. The power up to 10 mW is reported to be achievable [30] using the most popular two-photon ladder excitation to $5^2D_{5/2}$ state via $5^2P_{3/2}$ state that could be done owing to a combination of amplified spontaneous emission and parametric four-wave mixing. We decided to employ a similar excitation scheme but using $5^2D_{3/2}$ state of rubidium atom as the highest excited state. In this configuration more parametric processes are possible (see Fig. 1*c,d*).

In the first experiment we studied the properties of the blue light generated via the $5^2D_{3/2}$ intermediate state. Our setup was similar to the one reported in [31]. Two beams of light from the 776 nm ECDL and from the 780 nm distributed feedback laser diode (DFB) with a total power of over 150 mW were combined on a non-polarizing beam splitter and focused into a hot (120°C) rubidium cell without any buffer gas. Polarizations of two beams were controlled separately. The co-propagating geometry is crucial for the phase-matching condition to be satisfied.

An auxiliary rubidium vapour cell served as a reference for the 780 nm light, while the exact frequency of 776 nm light was determined with a commercial wavelength meter.

After the process of four-wave mixing had taken place in the hot rubidium cell, the beam of blue light generated in this cell was separated from the pump beams using a dichroic mirror and a colour filter. The intensity of the blue light was measured with a photodiode.

As could be predicted from the level scheme (Fig. 1*c*), we observed two blue wavelengths generated simultaneously, which was due to fine splitting of $6^2P$ orbital. Cascade decay through $6^2P_{1/2}$ state resulted in blue light having a wavelength of 421.6 nm, while 420.3 nm light was generated when atoms decayed through $6^2P_{3/2}$ state. The power of blue light was significantly lower than the one achieved when pumping to $5^2D_{5/2}$ state, due to an order in magnitude difference in the dipole moments, but the beam of blue light still seemed significantly brighter to the eye than the near-infrared pump beams.

We measured the combined intensity of light generated at two blue wavelengths as a function of pump lasers' detunings. The results are demonstrated in Fig. 5.

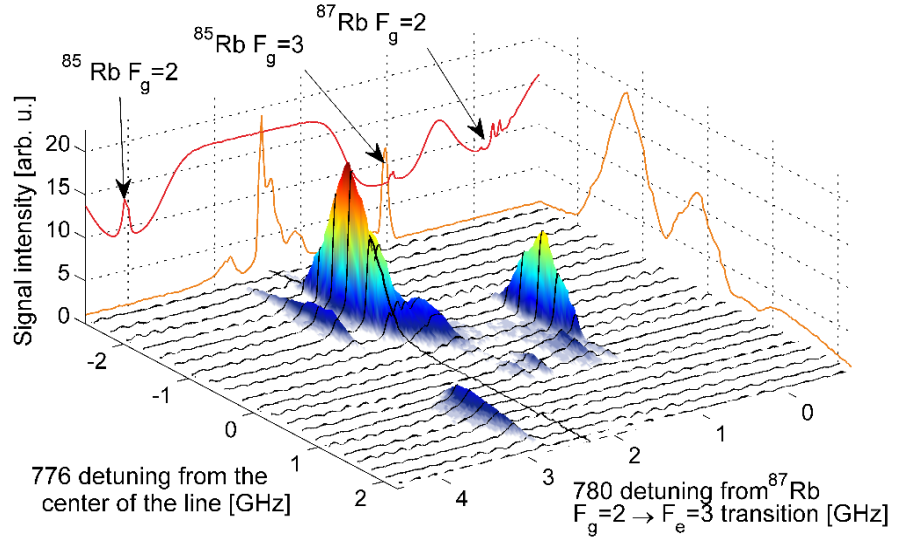

*Fig. 5.* The total intensity of generated blue light in cascade decay from $5^2D_{3/2}$ state as a function of 780 nm and 776 nm light detunings. Orange curves represent integrals over appropriate axes, while the red curve is a reference spectrum collected with a saturated absorption spectroscopy setup.

Similarly as in [31], we observed two dominant peaks. The resonant one, accompanied by strong blue fluorescence, occurred when the 780 nm light was tuned to F=3 → F'=2,3,4 transition group of $^{85}$Rb D2 line. The stronger, off-resonant one, took place when the 780 nm laser was blue-detuned from the above transition by 1.5 GHz. In the latter case the fluorescence is much lower than in the former. Both resonances occurred when lasers were tuned to the two-photon resonance.

The optimal polarizations of pump beams for blue light generation with $5^2D_{3/2}$ state proved to be significantly different from the ones that are optimal when pumping to the $5^2D_{5/2}$ state. In our case the polarizations should be linear and orthogonal to each other, while in the latter case the polarizations were shown to be optimal when they were equal and circular [31].

## 6. NON-DEGENERATE FOUR-WAVE MIXING

In the second experiment we made use of the same highest excited state to generate a beam of light at 762 nm in a pure, fully non-degenerate, parametric four-wave mixing process, in a so-called diamond configuration [32], as depicted in Fig. 1*d*. The same configuration has been used to generate entangled photon pairs from cold rubidium ensemble [33].

Beams from three lasers – the two used in the experiment described above and another DFB laser diode at 795 nm – were expanded to millimetre sizes and sent to the rubidium cell. The geometry was chosen in such a way that the phase-matching condition could be satisfied (see Fig. 6).

The fourth beam of light at 762 nm was generated in the cell, and then all four beams were focused. A multimode (200 μm core) fibre tip was located in the focal plane. We adjusted its position so that only the 762 nm light was well-coupled. The collected light was sent to a grating spectrometer. When detunings of all lasers ($\omega_{780}$, $\omega_{776}$, $\omega_{795}$) were adjusted in such a way that all the frequencies could add up to zero (i.e. $\omega_{780} + \omega_{776} - \omega_{795} - \omega_{762} = 0$, where $\omega_{762}$ is the exact resonant frequency of $5^2P_{1/2} \rightarrow 5^2D_{3/2}$ transition) we observed a narrow resonance of the 762 nm light generation.

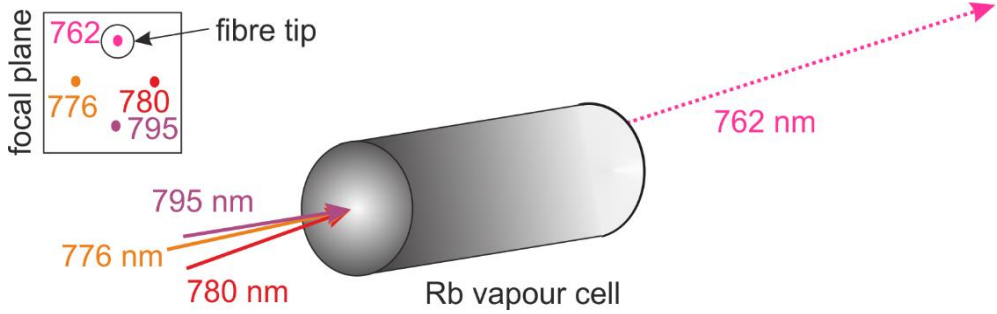

*Fig. 6.* Geometry for the four-wave mixing experiment in the diamond configuration of atomic levels.

Four-wave mixing in this configuration was possible using all ground state hyperfine split levels, of both [85]Rb and [87]Rb isotopes. We observed the intensities of the order of 100 pW, however optimizations of the entire setup are yet to be carried out.

## 7. CONCLUSIONS

The conclusions based on the results of our work are as follows.

The described theory of Raman scattering in rubidium vapour using a three-level atom model in the Λ-configuration admits treating this system as a two-level atom widely known in quantum optics [13].

The dynamics of the process of measuring the efficiency of photon generation as a function of pulse duration and writing laser power can easily be controlled by changing experimental parameters. This might be useful in future experiments with rubidium vapours.

The shielding from external magnetic field measured in our proof-of-principle experiment increases the amount of light registered on the camera during scattering process and increases by 30% the speed of signal growth.

The reconstruction of anti-Stokes scattering temporal dynamics makes it possible to characterize the rate of decay of atomic excitation.

The usage of a rich level structure of the rubidium atom opens a path to many new spontaneous and parametric processes. This has been illustrated by two examples, while many interesting alternatives are possible.

The coherent blue light generation as a parametric process [34] arises from a combination of four-wave mixing and amplified spontaneous emission, while the 762 nm light generation is a pure four-wave mixing process. These two experiments are first steps in designing a quantum interface between light and atomic ensembles [21] based on multi-photon processes, in principle similar to the one based on Raman scattering in Λ-configuration.